\documentclass{aa}  

\usepackage[switch]{lineno}
\usepackage{graphicx}
\usepackage{comment}

\usepackage{txfonts}
\usepackage{color}
\usepackage{comment}
\usepackage{soul}
\usepackage{longtable}
\usepackage{appendix}

\begin{document}

\title{Measuring the initial mass of $^{44}$Ti in SN 1987A through the $^{44}$Sc emission line}

   \author{R. Giuffrida \inst{1,2,3}, M. Miceli \inst{1,2}, E. Greco \inst{2}, S. Orlando \inst{2}, M. Ono \inst{4,5}, V. Sapienza \inst{2}, F. Bocchino \inst{2}, O. Petruk\inst{2,6}, B. Olmi \inst{7}, S. Nagataki \inst{8,9,10}}

\institute{Dipartimento di Fisica e Chimica E. Segr\`e, Universit\`a degli Studi di Palermo, Via Archirafi 36, 90123, Palermo, Italy
\and
INAF-Osservatorio Astronomico di Palermo, Piazza del Parlamento 1, 90134, Palermo, Italy
\and
AIM, CEA, CNRS, Université Paris-Saclay, Université de Paris, F-91191 Gif sur Yvette, France
\and
Institute of Astronomy and Astrophysics, Academia Sinica, Taipei 106319, Taiwan
\and
Astrophysical Big Bang Laboratory (ABBL), RIKEN Cluster for Pioneering Research, 2-1 Hirosawa, Wako, Saitama 351-0198, Japan
\and
Institute for Applied Problems in Mechanics and Mathematics, Naukova Street 3-b, 79060 Lviv, Ukraine
\and
INAF - Osservatorio Astrofisico di Arcetri, Largo E. Fermi 5, I-50125 Firenze, Italy
\and
RIKEN Center for Interdisciplinary Theoretical and Mathematical Sciences (iTHEMS), 2-1 Hirosawa, Wako, Saitama 351-0198, Japan
\and
Astrophysical Big Bang Laboratory (ABBL), RIKEN Pioneering Research Institute (PRI), 2-1 Hirosawa, Wako, Saitama 351-0198, Japan
\and
Astrophysical Big Bang Group (ABBG), Okinawa Institute of Science and Technology Graduate University (OIST), 1919-1 Tancha, Onna-son, Kunigami-gun, Okinawa 904-0495, Japan
}

\abstract
   {Deriving the mass and large-scale asymmetries of radioactive isotopes offers valuable insights into the complex phases of a supernova explosion. Important examples are $^{56}$Ni, with its decay products $^{56}$Co and $^{56}$Fe, and $^{44}$Ti, which are studied through their X-rays emission lines and provide a powerful diagnostic tool to probe the explosive nucleosynthesis processes in the inner layers of the exploding star.}
   {In this framework, SN 1987A provides a privileged laboratory being the youngest supernova remnant from which the mass of Ti has been estimated. However, some tension exists in determining the initial mass of $^{44}$Ti. Previous analysis, relying on \textit{NuSTAR} and \textit{INTEGRAL} data, report $M_{44} = (1.5 \pm 0.3) \times 10^{-4}$ $M_\odot$ and $M_{44}=(3.1 \pm 0.8) \times 10^{-4} M_\odot$, respectively. In this paper we estimate the initial mass of $^{44}$Ti via its decay product, the $^{44}$Sc emission line at 4.09 keV, using \textit{Chandra} observations.}
   {We perform multi-epoch spectral analysis focusing on the inner part of the remnant, to minimize the contamination from the X-ray emission stemming from the shocked plasma. As a result, we provide the detection of $^{44}$Sc emission line in the central part of SN 1987A with a $\sim$99.7\% (3 $\sigma$) significance.}
   {The simultaneous fit of the spectra extracted from observations between 2016 and 2021 provides a line flux of $6.8^{+2.2}_{-2.3}\times 10^{-7}$ photons s$^{-1}$ cm$^{-2}$ corresponding to a $^{44}$Ti mass $M_{44}=(1.6\pm0.5) \times 10^{-4} M_\odot$ (errors at the $90\%$ confidence level). The results obtained with our spectral analysis seem to align with those derived with NuSTAR.}{}

\keywords{ISM: supernova remnants - X-rays: individuals: SN 1987A - Nuclear reactions, nucleosynthesis, abundances}

   \titlerunning{Measuring 44Ti mass in SN 1987A}
   \authorrunning{Giuffrida et al.}

   \maketitle

\section{Introduction} 
\label{sec:intro}

Supernova (SN) explosions are important sources to study the chemical evolution of the Universe.  The supernova ejecta carry information on the explosive nucleosynthesis processes, and elements synthesized in the inner layers of core-collapse supernvoae can “keep memory” of the physical mechanisms governing the explosion. 
Important issues can be addressed by studying the radioactive emission of the $^{56}$Ni and $^{44}$Ti isotopes, which are synthesized in the central part of the exploding star \citep{Hashimoto1995,nhs97,Nagataki2000}, together with their daughter products, such as $^{56}$Co and $^{56}$Fe for $^{56}$Ni, $^{44}$Sc for $^{44}$Ti. In particular, the yield of $^{44}$Ti is very sensitive to the supernova shock conditions (much more than $^{56}$Ni), namely peak temperature and density of the ejecta reached soon after the core collapse  (e.g., \citealt{Magkotsios10}), thus providing a powerful diagnostic tool for the explosion physics. Moreover, as it has been found in \cite{nhs97,nmk98,Nagataki2000}, a high ratio $^{44}$Ti/$^{56}$Ni can be considered as a signature of of large-scale anisotropies in the explosions.

After the complete decay of $^{56}$Co and $^{57}$Co, which dominate the energy balance for the first few years after the explosion, the IR, optical and UV emissions are driven by $^{44}$Ti radioactive decay.
Additional evidence for the presence of $^{44}$Ti can be derived from the X-ray and $\gamma$-ray emission of its decay chain: $^{44}$Ti decays in $^{44}$Sc with an e-folding time of 85 years \citep{agm06}, producing two emission lines at 67.9 keV and 78.4 keV through electron capture. Subsequently $^{44}$Sc decays in $^{44}$Ca in a lifetime of 5.7 hours \citep{agm06} emitting a line at 1157.0 keV through $\beta$-decay and electron capture. The electron capture results in a $^{44}$Sc with a vacancy in its K-shell, which produces a K-shell transition with an X-ray emission line at 4.09 keV. 

The supernova SN 1987A, observed on February 23 1987  in the Large Magellanic Cloud \citep{West1} at a distance $d = 51.4$ kpc \citep{Panagia1999}, is an ideal target to study nearby, very young, core-collapse supernova (CCSN) explosion and its subsequent evolution into a supernova remnant (SNR).  
X-ray emission from SN 1987A has been detected by \cite{dhi87} in the hard band (above 10 keV) a few months after the explosion, and, a few years later, by \citet{bbp94} with \textit{ROSAT} in the soft X-rays. Afterwards, starting from 1999, multi-epoch \textit{Chandra} observations have revealed the morphology of the X-ray emission and its time evolution  (\citealt{hbd13,fzp16,rpz24}, see also Fig \ref{fig:img}). The X-ray emission of SN 1987A is characterized by a ring-like structure in expansion, which is the result of shocked circumstellar medium produced before the SN explosion by stellar winds from the progenitor star and previously observed in the optical band (\citealt{lm91L,lf91,bhf95,cd95}). Important information on the origin and evolution of the X-ray spectral features have been obtained thanks to the analysis of \textit{XMM-Newton} observations \citep{hga06,svc21} and \textit{SRG/eRosita} \citep{mhs22}.

The X-ray emission can be well reproduced by
magnetohydrodynamic (MHD) simulations \citep{omp15,omp19,oon20, omo25} that self-consistently describe the time evolution of the remnant morphology, fluxes, and spectra. Moreover, the comparison between X-ray data and MHD simulations provide important information to ascertain the origin of the broadening of X-ray emission lines, pinpointing the role of thermal broadening and Doppler broadening associated with the expansion of the plasma  \citep{mob19,rpz21,miceli23}. Recently, \citet{katsu25} have shown that with \textit{XRISM} observations it is possible to recover the expansion velocity of the outer (i.e., metal-poor) ejecta layers interacting with the reverse shock, as predicted by \citet{smb24,smb24b}.Thermal X-ray emission from metal-rich ejecta is expected to be modest only in the next few years \citep{omo25}. 

Multi-wavelength observations of SN 1987A have revealed Doppler shifts in emission lines of heavy elements (e.g. [Fe II] and [Ni II]) up to velocities $>$ 3000 km s$^{-1}$ (e.g. \citealt{hcs90,che94,uca95,lfs16}).
The formation of $^{56}$Co gamma-ray lines in SN 1987A several months earlier than expected \citep{msl88}, reveals a mix between some innermost heavy nuclear products and the outer envelope. 
The $^{56}$Co lines at 847 and 1238 keV have been detected by balloon experiments \citep{cpp88,mvj88,snc88,tbg89} and by the $\gamma$-ray spectrometer (GRS) on board the Solar Maximum Mission (SMM) satellite \citep{msl88,ls90}.
As a result, it has been found that almost 5\% of iron-group rich plasma mixed out to velocities of $\sim$ 3000 km s$^{-1}$ \citep{abk89,ls90}, much higher than the characteristic values ($\la 1000$ km s$^{-1}$) expected for iron-group material. This mixing can be due to either Rayleigh-Taylor instabilities or to radioactive heating causing the expansion of $^{56}$Ni/$^{56}$Co rich bubbles \citep{basko94,kpj03,utu18,onf20}.

The anisotropy in the inner ejecta velocities has been confirmed with the detection of a redshift of about 0.23 keV in the  $^{44}$Ti X-ray emission lines at 67.87 keV \citep{bhm15}, observed with \emph{NuSTAR} and corresponding to a velocity of $\sim$ 700 km s$^{-1}$ (in the rest frame of SN 1987A). \citet{bhm15} measured the flux of the line at 67.87 keV ($F_{68} = 3.5 \pm 0.7 \times 10^{-6}$ cm$^{-2}$ s$^{-1}$, similar results were later obtained by \citealt{alj21}), thus deriving an initial mass  of $^{44}$Ti $M_{44}=1.5 \pm 0.3 \times 10^{-4} M_{\odot}$ (see Eq.\ref{eq:flux}). This value  is in agreement with simulations \citep{thn90,wh91}, which predict $M_{44}=0.2 - 2.5\times 10^{-4} M_{\odot}$ in CCSNe. 
The modeling of the optical spectrum performed by \cite{jfk11} also suggests $M_{44}=1.5\pm0.5\times 10^{-4} M_{\odot}$. Moreover the model B18.3 developed in \cite{oon20,onf20} predicted and initial mass of $^{44}$Ti of about $1.4 \times 10^{-4} M_\odot$.
However, this value should be taken with some cautions since the explosive nucleosynthesis is taken into account with a small approximate nuclear reaction network and this may determine an overestimation of the $^{44}$Ti abundance \citep{mon15}.

On the other hand, the fluxes of the $67.87$ keV and $78.32$ keV lines measured with \textit{INTEGRAL} indicate a value for the initial amount of $^{44}$Ti higher by a factor of $\sim$ 2, i.e. $M_{44} = (3.1 \pm 0.8) \times 10^{-4} M_\odot$ \citep{glt12}.

The $^{44}$Ti yields can also be estimated by measuring the flux of the $^{44}$Sc line at 4.09 keV, as successfully done for the supernova remnant G1.9+0.3 by \citet{brg10}, thanks to the analysis of \textit{Chandra} observations. This approach has been pursued for SN 1987A by \citet{Leising06}, who analyzed multi-epoch \emph{Chandra} spectra of the entire remnant finding no significant detection of the $^{44}$Sc line due to high contamination by the emission from the shocked ring material. Their upper limit on the line flux corresponds to $M_{44}<2 \times 10^{-4} M_\odot$, without including the effects of the photoelectric absorption from the surrounding cold ejecta.

In this paper we adopt a new methodology to detect the $^{44}$Sc line in multi-epoch \emph{Chandra} observations of SN 1987A and estimate the value of $M_{44}$. In particular, we select a circular region in the inner part of the remnant (with angular radius $R\sim0.3''$, see Fig. \ref{fig:img}), where the contamination of thermal emission from shocked plasma is as low as possible.
Additionally, the expansion of the bright ring reduces contamination of the thermal emission from shocked plasma in the central region (where $^{44}$Sc is supposed to be synthesized) at the latest epochs. We also note that the lowest absorption from cold ejecta at the latest epochs favors the detection of the $^{44}$Sc emission line. 

The paper is organized as follows: Sec. \ref{sec:reduction} presents the observations analyzed and the data reduction; our results are shown in Sec. \ref{sec:results}, \ref{sec:xrism}  and, finally, the discussion and conclusions are drawn in Sec. \ref{sec:discussion}. All the information about the observations are listed in Appendix \ref{app:obs}, Appendix \ref{evolution} shows the temporal evolution of the $^{44}$Sc line, the background (region and spectrum) is presented in Appendix \ref{app:bkg}, while Appendix \ref{app:corner} shows the complete view of the Markov chain Monte Carlo (MCMC) corner plot discussed in Sect. \ref{sec:results}.

\section{Observations and Data Reduction}
\label{sec:reduction}
We analyzed all the archive \textit{Chandra}/ACIS-S observations of SN 1987A, spanning 22 years from 2000 to 2021. The relevant information about the observations are summarized in Table \ref{tab:obs}. 
Data were analyzed with \textit{CIAO} (v4.13), using \textit{CALDB} (v4.9.4), and reprocessed with the \textit{chandra\_repro} task. 

For the image analysis, we adopted the subpixel sampling with a pixel size of 0.06 arcsec, for each observation. We improved the spatial resolution of all of the images with the \textit{CIAO} \texttt{arestore} tool, which takes into account the Point Spread Function (PSF) of the telescope by using the Lucy-Richardson deconvolution algorithm (as it has been already done in \citealt{hbd13,fzp16,rpz24}). In order to produce the PSF for the deconvolution, we used the \texttt{MARX} software \citep{marx}. The deconvolved image better constrains the emission from the ring, providing a good starting point to select the region used for the spectral analysis.

Spectra have been extracted from the event files using \textit{specextract}, which also generates the corresponding ancillary response file (ARF) and redistribution matrix (RMF), and rebinned with the optimal binning algorithm \citep{kb16}. Spectra extracted from observations performed within one month have been merged using the the \textit{CIAO} tool \texttt{combine\_spectra}.

Spectra were analyzed using the software \textit{XSPEC} V 12.13.0c \citep{arn1996}, in the 1.0-5.5 keV band. 
Because of the low statistics, we fitted the spectra by following the C-statistic \citep{Cash79}. 
To calculate the error bars, we run the MCMC algorithm within \texttt{xspec}, thus accounting for the dependencies among the free parameters in the fit. We used the Goodman-Weare algorithm with a chain length of $2 \times 10^5$ and 30 walkers. 

\section{Results}
\label{sec:results}

$^{44}$Ti is synthesized in the innermost ejecta, and its X-ray emission can become detectable only when the gas becomes optically thin, i. e., about $20$ years after the SN explosion (e.g., \citealt{fc1987}).
Once the ejecta become transparent to  the X-rays, the flux corresponding to each radioactive emission line, $F_i$, can be calculated as
    \begin{equation}
    \centering
    F_i = \frac{ M_{44} W_i }{ 4 \pi d^2 44m_p t_{44} }e^{-t/t_{44}}
    \label{eq:flux}
    \end{equation}
where $M_{44}$ is the initial mass of $^{44}$Ti, \textit{d} is the distance to the source, $m_p$ is the proton mass, $t_{44}$ is the e-folding time decay of $^{44}$Ti ($\sim 85$ yr) and $W_i$ is the emission efficiency for the three X-ray emission lines (17.4\% for line at 4.1 keV, 87.7\% for line at 67.87 keV and 94.7\% for line at 78.4 keV, \citealt{glt12}). 

\begin{figure}
\centering
\includegraphics[width=\columnwidth]{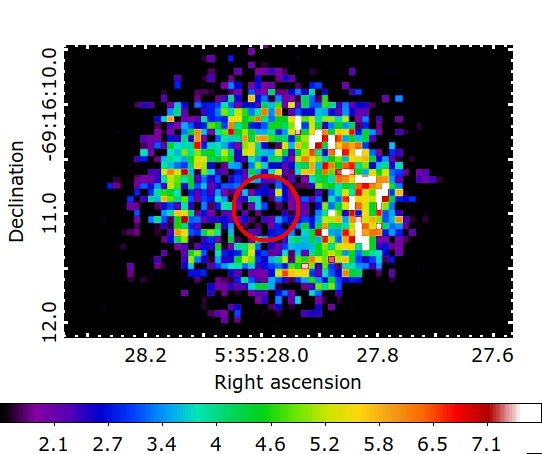} \\
\includegraphics[width=\columnwidth]{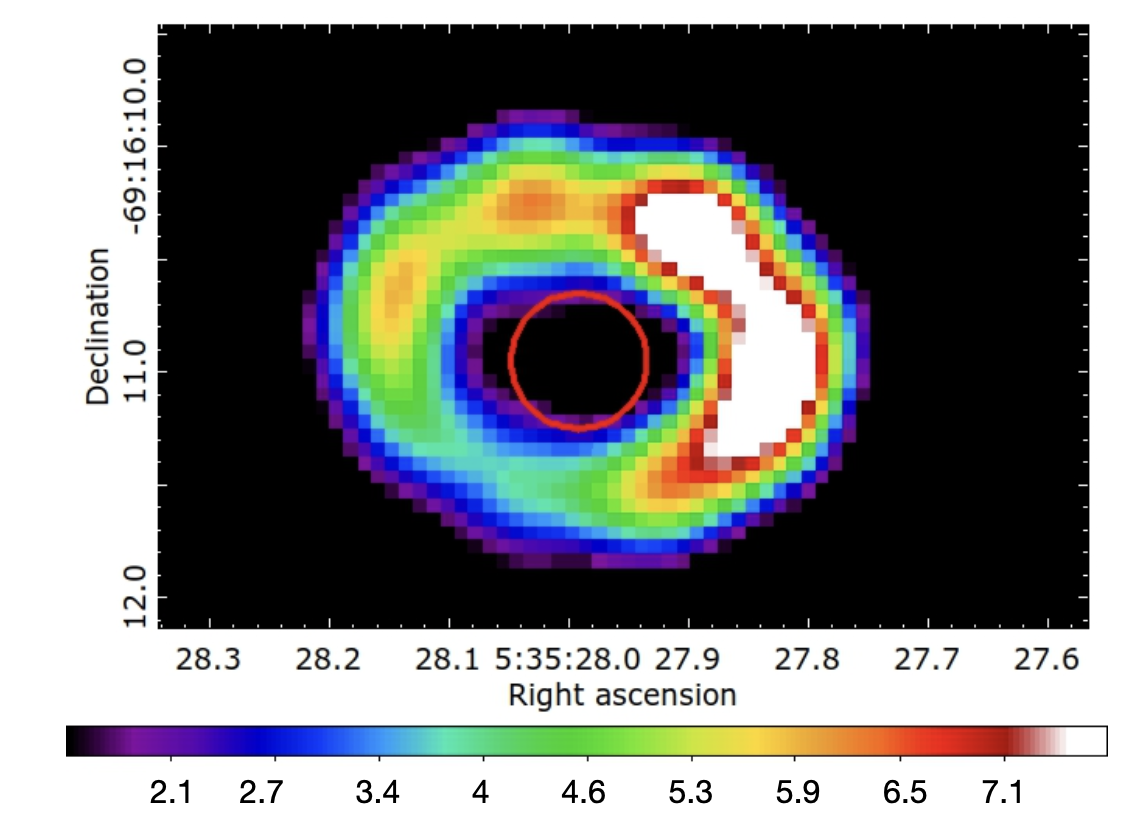}
	\caption{\emph{Upper panel:} \emph{Chandra/}ACIS photon count map of SN 1987A in 2019 (Obs ID 21304) in the 0.5-7.0 keV energy band. \emph{Lower panel:} same image as above, but deconvolved using the Lucy-Richardson algorithm. Both images have a pixel size of 0.06". The red circle indicates the region selected for spectral analysis.}
	\label{fig:img}
\end{figure}
We expect the $^{44}$Ti in SN 1987A to be distributed over an extended area, somehow similar to that of Fe-rich ejecta. \citet{lfs23} showed that the Fe-rich ejecta are located within the main shell of reverse-shocked material in a region denominated the homunculus, well inside the shocked dense equatorial ring (ER). Accordingly, we select a circular region with radius $0.3''$ (shown as a red circle in Fig. \ref{fig:img}) for spectral extraction. Within this region, we expect the maximum $^{44}$Sc line emission from the unshocked ejecta, while, at the same time, the choice of the extraction region minimizes the contamination associated with the thermal X-ray emission from the ER.

The source spectrum has been fitted with a model taking into account the interstellar absorption (\texttt{Tbabs} model) with a column density N$_{\rm{H}}$ fixed at $2.35\times10^{21}$ cm$^{-2}$ \citep{pbg06}, one non-equilibrium of ionization collisional plasma model (\texttt{vnei} model), to reproduce thermal emission from the shocked plasma spilling from the bright ring inside the central region, because of the telescope PSF, an absorbed (the absorption originating from cold ejecta) power law to account for the nonthermal emission stemming from the putative pulsar wind nebula  \citep{gmo21,gmo22} (\texttt{vphabs*pow} model) and a Gaussian component for the $^{44}$Sc line. The $^{44}$Sc line emission is expected to be redshifted as that of $^{44}$Ti ($(\Delta E/E)_{^{44}\rm{Ti}} = (\Delta E/E)_{^{44}\rm{Sc}}$).
Since the emission line at $67.87$ keV shows $(\Delta E)_{^{44}\rm{Ti}}= 0.23 \pm 0.09$ \citep{bhm15}, we expect $(\Delta E)_{^{44}Sc}=0.013\pm 0.005$ keV, obtaining the line centered at about 4.076 keV (we kept the line centroid frozen to this value in the fitting process).
The model for the source spectra is then described by the following equation:
\begin{equation}
    \textit{Src mod} = \textit{Tbabs} \times (vphabs\times pow + vnei + gauss)
    \label{eq:source}
\end{equation}

In the early \emph{Chandra} observations we expect the absorption from cold ejecta to be prominent at the energy of the $^{44}$Sc emission line and the meaurement of the line flux to be affected by this issue. Indeed, a careful search for a line emission at 4.076 keV provides very low line fluxes for the observations performed between 2000 and 2015, as described in detail in Appendix \ref{evolution}. The detailed modeling of the (time-dependent) absorption from cold ejecta is beyond the scope of this paper. We also notice that, given the uncertainties in the distribution of $^{44}$Ti in our simulations of SN 1987A (see \citealt{onf20}), it is not possible to perform an accurate estimate of the emission and absorption of the line in the ejecta. However, we warn the reader that this can somehow affects our estimates.
We here focus on the spectral analysis of the latest 6 years of our set of observations (2016-2021), when the expansion of the remnant significantly reduces the thermal contamination in the central region and the cold ejecta are expected to be optically thin, allowing Eq. \ref{eq:flux} to be applied.

To improve the statistics in our study of the $^{44}$Sc emission line, we simultaneously fitted spectra extracted from all the observations performed between 2016 and 2021. When fitting the spectra, the electron temperature, ionization parameter and normalization of the \texttt{vnei} component of each spectrum are left free to vary. The normalization of the Gaussian component modeling the $^{44}$Sc line is also a free parameter, but, in this case, normalizations from different epochs are tied together, accounting for the expected exponential time decay of Eq. \ref{eq:flux}. The sigma of the Gaussian component has been frozen to the value $1 \times 10^{-3}$ keV in order to reduce the number of free parameters. We verified that the results of the fits are not affected by letting it free to vary.

\begin{figure}[h!]
\includegraphics[width=\columnwidth]{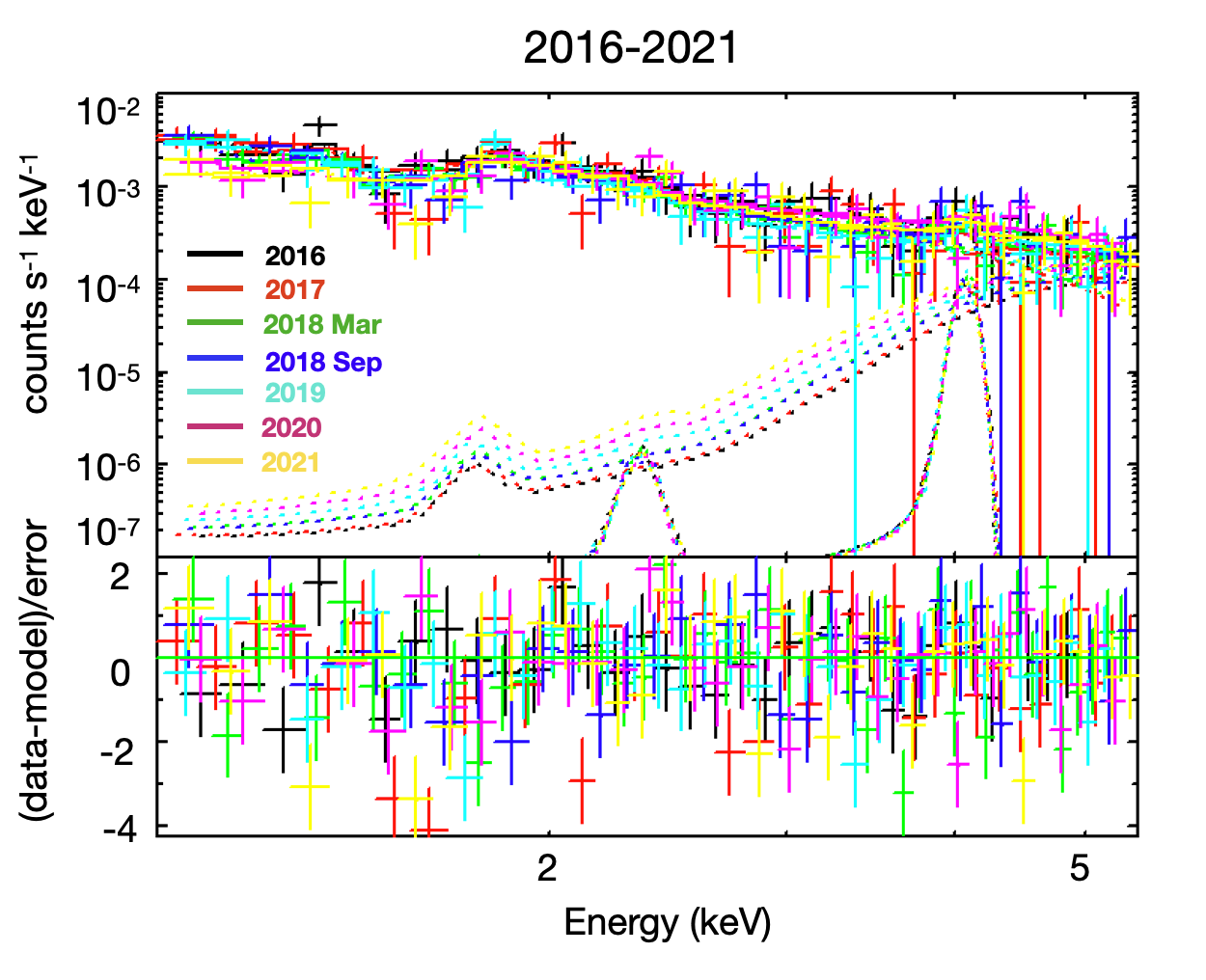}
\includegraphics[width=\columnwidth]{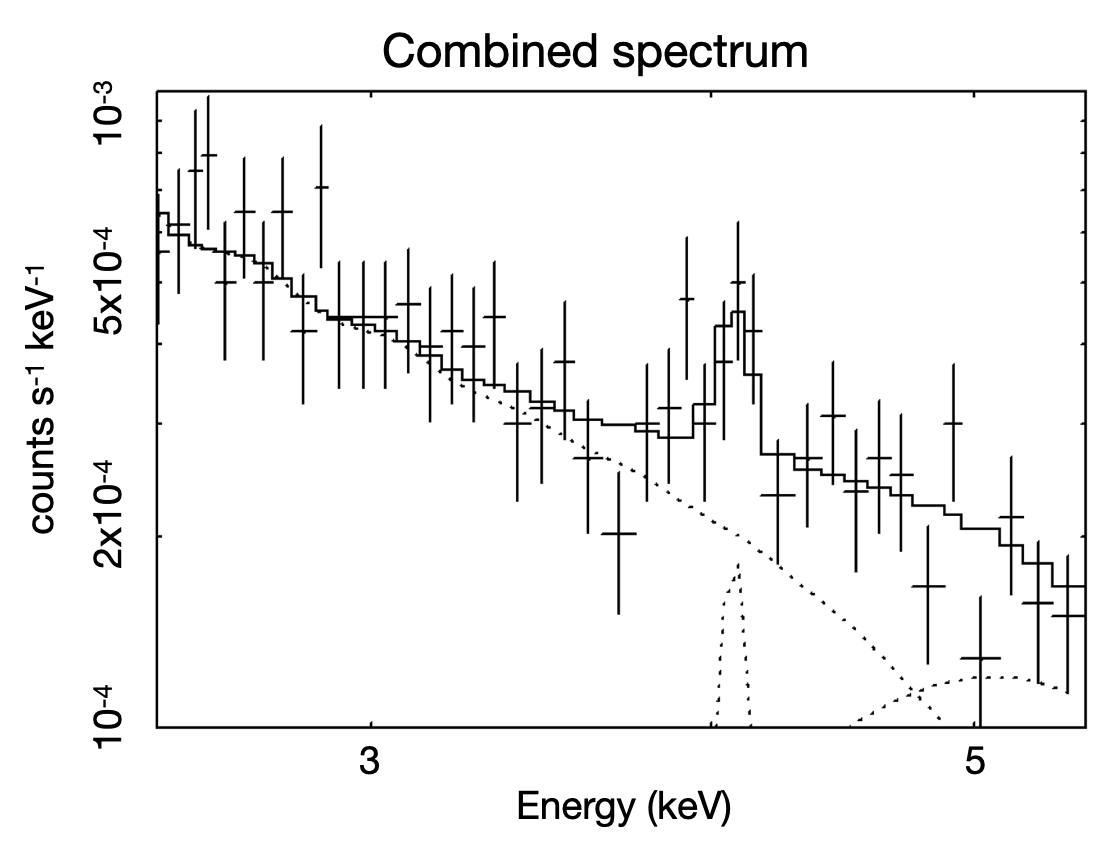}
	\caption{\emph{Upper panel:} \emph{Chandra}/ACIS spectra extracted from the circular region shown in Fig. \ref{fig:img} for all observations performed between 2016 and 2021 (See Tab. \ref{tab:obs}), with the corresponding best-fit model and residuals. Dotted lines show every component of the best fit model for each epochs. \emph{Lower panel}: combined spectrum obtained by summing all the observations in the upper panel. The $^{44}$Sc line is modelled by the narrow Gaussian at $4.076$ keV. Dotted lines show the different components of the best fit model.}
	\label{fig:spectrum_2016-2021}
\end{figure}

The background region has been chosen in the same chip as the source region, in an area without visible point-like sources, and the corresponding spectrum has been fitted with an ad hoc model (see Appendix \ref{app:bkg} for details).
The background flux is always less than 0.1\% of the total (this is the reason why the background components are not visible in the spectral plots).
Data with the corresponding best fit model and residuals are shown in the upper panel of Fig. \ref{fig:spectrum_2016-2021}

We notice that the angular extension of the spectral extraction region shown in Fig. \ref{fig:img} ($0.3''$) is lower than that of the \emph{Chandra} PSF ($\sim0.5''$), so part of the photons emitted within the central region are scattered outside from it by the telescope mirrors. 
\citet{lfs23} showed that the unshocked Fe-rich ejecta in SN 1987A extend over a large area inside the ER, somehow filling entirely our spectral extraction region. As an educated guess, we assume that Ti-rich ejecta are uniformly distributed within the extraction region \citep{onf20}. Under this hypothesis, the $^{44}$Sc line emission originates from a disk with radius $0.3''$.
By adopting the \texttt{MARX} software, we find that, in this scenario, $3/4$ of the photons are scattered outside from the region. A very similar percentage (i.e., $71\%$) is obtained by assuming for the emission a 2-D Gaussian profile with $\sigma=0.15''$. 
Though we warn the reader that this percentage might slightly vary for different spatial distributions of the surface brightness within the extraction region, we expect that the line flux obtained with the spectral analysis needs to be multiplied by a factor $f$ of the order of $f\approx4$ to recover its intrinsic value. We will adopt $f=4$ hereafter.

\begin{figure}[ht!]
\includegraphics[width=\columnwidth]{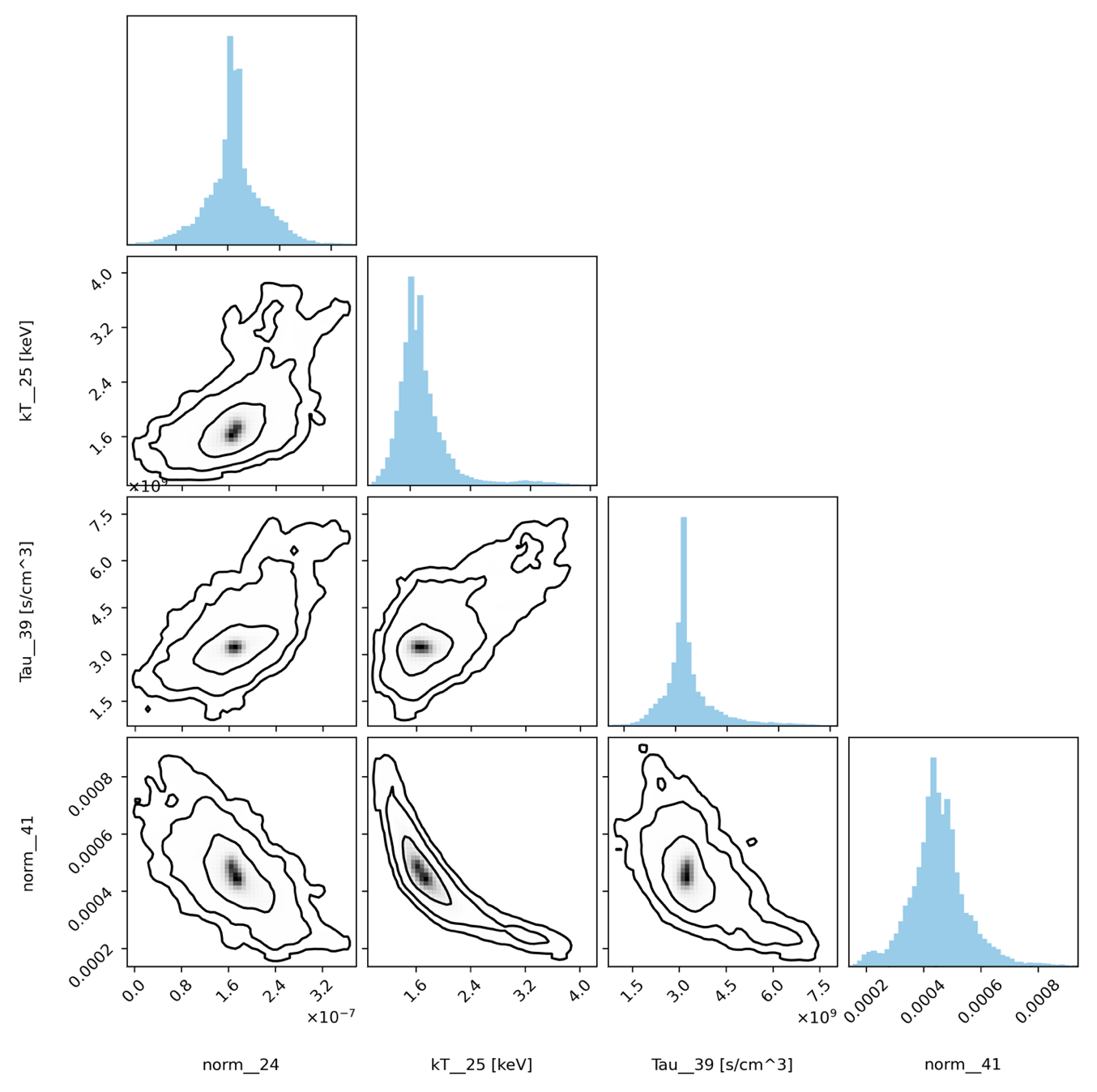}
	\caption{Close-up view of the MCMC corner plot shown in Fig. \ref{fig:corner_plot_global} (Appendix \ref{app:corner}). The parameter \texttt{norm\_{24}} indicates the normalization of the Gaussian at 4.076 keV, not corrected for the PSF effects (see text). \texttt{kT\_{25}, Tau\_{39}, norm\_{41}} are the temperature, ionization parameter and normalization for the \texttt{vnei} component in the 2016 spectrum, respectively.}
	\label{fig:corner_plot}
\end{figure}

From the combined analysis of the 2016-2021 spectra, we find a line flux of $6.8^{+2.2}_{-2.3}\times 10^{-7}$ photons s$^{-1}$ cm$^{-2}$ (error bars at the $90\%$ confidence level). For visualization purpose, lower panel of Fig. \ref{fig:spectrum_2016-2021} shows the combined \emph{Chandra}/ACIS spectrum, obtained by summing all the spectra collected between 2016 and 2021: the $^{44}$Sc line is clearly visible above the continuum.  Fig. \ref{fig:corner_plot} shows a detail of the MCMC corner plot for the first 4 free parameters, namely the normalization of the Gaussian centered at 4.076 keV, the electron temperature $kT$, the ionization age $\tau$ and the normalization of the first \texttt{vnei} component associated with the spectrum observed in 2016. Contours are at 68\%, 95.5\% and 99.7\% confidence levels. The corresponding complete corner plot is shown in Fig. \ref{fig:corner_plot_global} (Appendix \ref{app:corner}) revealing the significance of the detection at about 3$\sigma$ (99.7\% confidence level). 
We conclude that the detection of the line is robust and statistically significant.

Following Eq. \ref{eq:flux}, setting $W_i$ to 0.174 (see Sect. \ref{sec:intro}), we can estimate the initial mass of $^{44}$Ti from the $^{44}$Sc line flux reported above, finding $M_{44}=1.6\pm0.5 \times 10^{-4}$ M$_\odot$ (error bars at the $90\%$ confidence level). Our results are in remarkably good agreement with those obtained from the analysis of \textit{NuSTAR} spectra by \cite{bhm15}. On the other hand, our estimate of the $^{44}$Ti initial mass is significantly lower than that reported by \cite{glt12} on the basis of  the analysis of \textit{INTEGRAL} data.

\begin{figure}[ht!]
    \centering
    \includegraphics[width=\columnwidth]{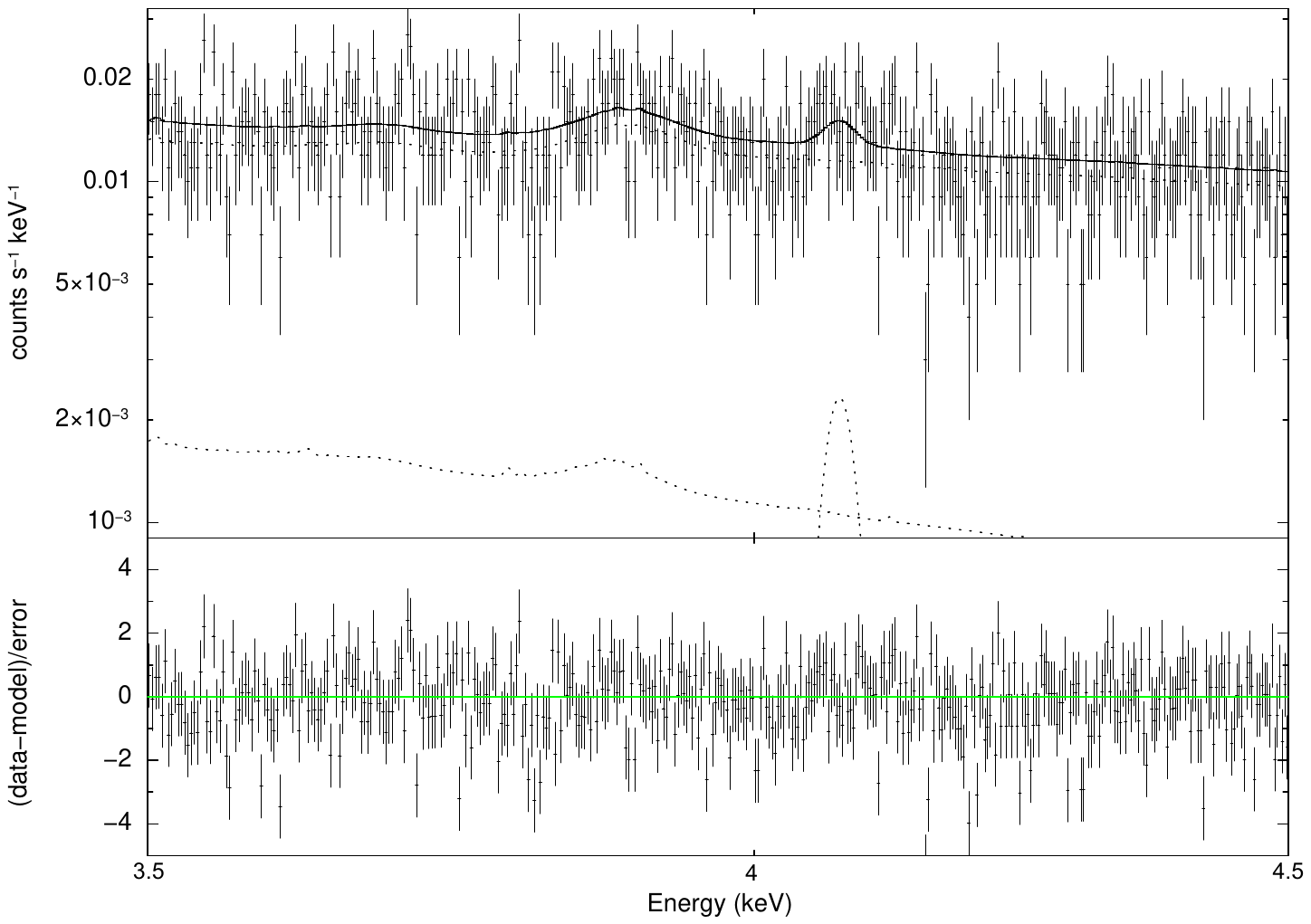}
    \includegraphics[width=\columnwidth]{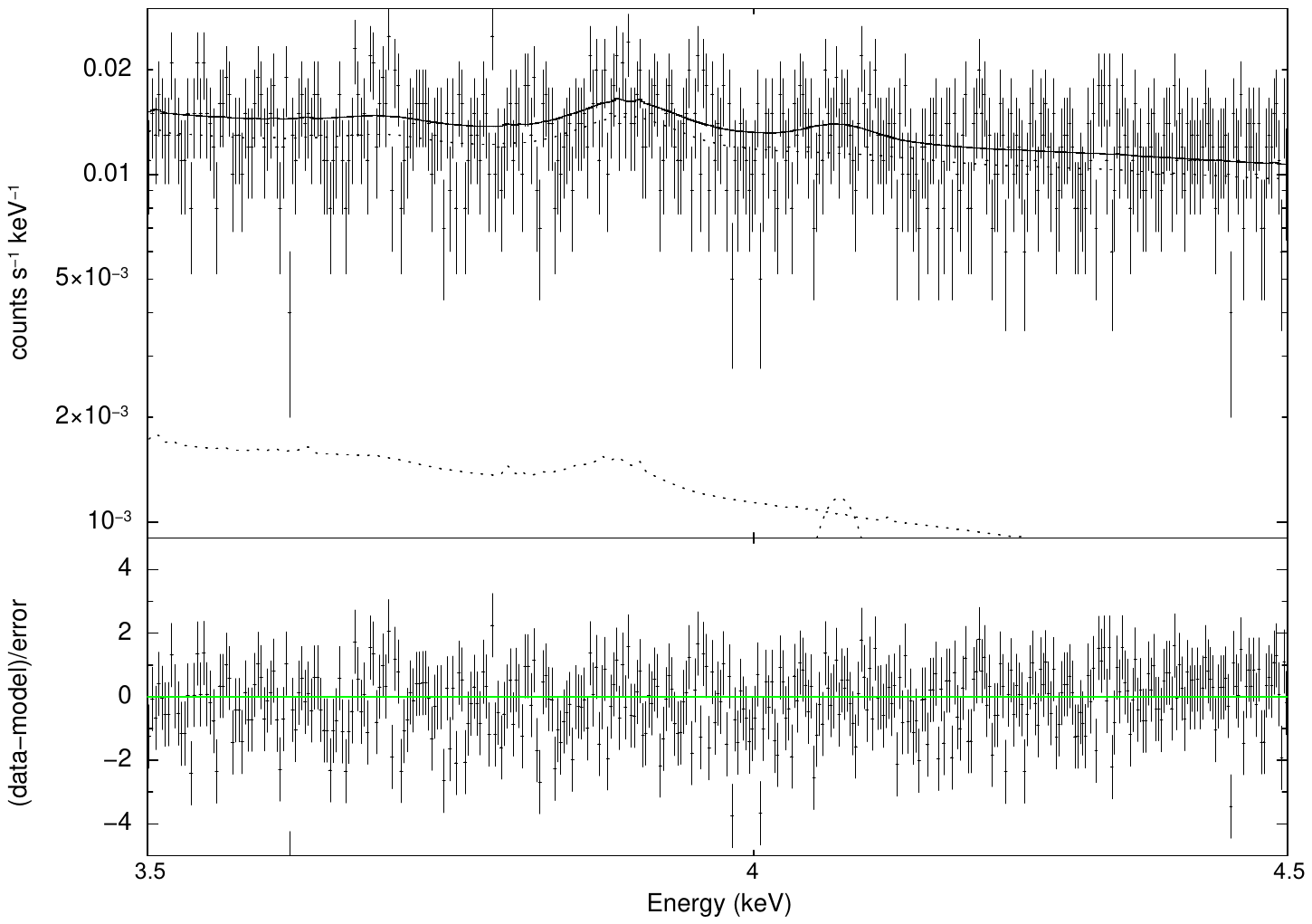}
    \caption{XRISM-Resolve synthetic spectra produced from the model by \cite{smb24} with the addition of the expected Scandium line for year 2025, with updated RMF and ARF including the effects of the gate valve closed \citep{smb24b}. Dotted line show the different components of the best fit model. The exposure time is 400 ks.
    \textit{Upper panel}: Spectrum synthesized including the best-fit Gaussian component with a line width corresponding to a Doppler broadening of 1000 km s$^{-1}$.
    \textit{Lower panel}: Same as upper panel but for a Doppler broadening of 2000 km s$^{-1}$. }
    \label{fig:xrismline}
\end{figure}

\section{\textit{XRISM-Resolve} simulated spectra}
\label{sec:xrism}
The detection of the $^{44}$Sc line might be possible with the new X-ray telescope, XRISM. SN 1987A will not be spatially resolved by the XRISM mirrors, so in this case it will not be possible to extract the spectra from small regions to reduce the contamination from thermal X-ray emission as we have done with \emph{Chandra}. On the other hand, the high spectral resolution provided by the XRISM-Resolve spectrometer will help in making the line emerge over the continuum.

We simulated XRISM-Resolve spectra for the year 2025 leveraging a phenomenological model which reproduces the spectrum from \cite{smb24,smb24b}, including the effects of the Gate Valve closed. We assumed an exposure time of 400 ks, which is similar to the actual exposure time for SN 1987A in the XRSIM Performance Verification phase observation.
To this model we added a Gaussian to account for the  $^{44}$Sc line and an absorbed power law \citep{gmo21,gmo22} for the year 2024 to account for the emission of the putative PWN.
In Fig. \ref{fig:xrismline} we show the simulated spectra obtained by assuming two different $^{44}$Sc line widths (the line being Doppler-broadened because of the rapid expansion of the ejecta), namely 1000 km s$^{-1}$ and 2000 km s$^{-1}$ in the left and right panels respectively.

The detectability of the line strongly depends on its broadening. By considering an expansion velocity of 1000 km s$^{-1}$ the significance of the detection exceeds the $99\%$ confidence level. On the other hand, by assuming a much more reasonable value of $2000$ km s$^{-1}$ for the expansion velocity, the analysis of the synthetic spectrum shows a non-detection of the $^{44}$Sc line, its flux being larger than zero at only the 68\% confidence level.
Therefore, our results indicate a non-detection of the $^{44}$Sc line with XRISM-Resolve. Our conclusions nicely align with the recent findings by \cite{katsu25}.

\section{Discussion and Conclusion}
\label{sec:discussion}

The study of $^{44}$Ti in SNRs plays a crucial role in understanding the physical processes governing the explosion of massive stars. Previous works have already detected radioactive emission lines of $^{44}$Ti in SN 1987A (\citealt{bhm15,glt12}), though \emph{NuSTAR} and \emph{INTEGRAL} spectra provide different fluxes, thus making the initial mass of $^{44}$Ti still debated. In particular, using the relation between the flux and the initial mass of $^{44}$Ti, Eq. (\ref{eq:flux}), it has been found $M_{44} = (1.5 \pm 0.3) \times 10^{-4} M_{\odot}$ with \textit{NuSTAR} data analysis \citep{bhm15}, and $M_{44} = (3.1 \pm 0.8) \times 10^{-4} M_\odot$ with \textit{INTEGRAL} \citep{glt12}. Another method to measure the initial mass of $^{44}$Ti involves the emission line of $^{44}$Sc, which is a  product of the $^{44}$Ti decay chain.

In this work, we performed a systematic search for the $^{44}$Sc line by analyzing multi-epoch \emph{Chandra} observations of SN 1987A (from 2000 to 2021). We exploited the remarkable spatial resolution of the \textit{Chandra} mirrors to extract X-ray spectra from a small region (radius 0.3 arcsec) in the center of the remnant, as shown in Fig. \ref{fig:img}. 

While our quest for the $^{44}$Sc line is affected by the absorption from the surrounding cold ejecta at early epochs (as predicted by \citealt{fc1987}), we significantly detect the $^{44}$Sc line by analyzing the spectra extracted from observations performed between 2016 and 2021, when cold ejecta are expected to be optically thin. In particular, we measure a line flux $6.8^{+2.2}_{-2.3}\times 10^{-7}$ photons s$^{-1}$ cm$^{-2}$, corresponding to an initial mass $M_{44}=(1.6\pm0.5) \times 10^{-4} M_\odot$. To our knowledge, this is the first firm detection of the Sc emission line in SN 1987A.

The precise estimation of the mass of $^{44}$Ti depends on the actual size and morphology of the emitting region. As described in Sect. \ref{sec:results}, our analysis has been conducted assuming the $^{44}$Sc emission to stem from a circular region with radius $0.3''$. According to \cite{lfs23}, we consider this extent as an upper limit. As a results, also our initial mass of $^{44}$Ti must be considered as an upper limit value. Nevertheless, we showed that even by assuming a different distribution of the line surface brightness, our results do not change dramatically ($<10\%$). Our estimate of the $^{44}$Ti mass is in remarkable agreement with \textit{NuSTAR} analysis \citep{bhm15}. On the other hand, it is substantially lower than that obtained by \cite{glt12} through \textit{INTEGRAL} observations.

Our value of $M_{44}$ is also in good agreement with long-term 3-D simulations of neutrino-driven explosions \citep{skj23}, considering an explosion energy of $(1.3 - 1.5) \times 10^{51}$ erg \citep{Arnett87, Utrobin05, uwj21}.

New \emph{Chandra} observations will be able to increase the statistics and provide a more constrained estimate of $M_{44}$.

In conclusion, our work provides an independent procedure to measure the yields of $^{44}$Ti in SN 1987A through the analysis of the soft X-ray emission. Future observations will provide tighter constraints.

\begin{acknowledgements}
M.M., V.S., E.G., S.O., and F.B. acknowledge financial contribution from the PRIN 2022 (20224MNC5A) - ``Life, death and after-death of massive stars'' funded by European Union – Next Generation EU,  and the INAF Theory Grant ``Supernova remnants as probes for the structure and mass-loss history of the progenitor systems''.
This paper is partially supported by the  Fondazione  ICSC, Spoke 3 Astrophysics and Cosmos Observations. National Recovery and Resilience Plan (Piano Nazionale di Ripresa e Resilienza, PNRR) Project ID CN\_00000013 ``Italian Research Center on  High-Performance Computing, Big Data and Quantum Computing"  funded by MUR Missione 4 Componente 2 Investimento 1.4: Potenziamento strutture di ricerca e creazione di ``campioni nazionali di R\&S (M4C2-19 )" - Next Generation EU (NGEU).
R.G., V.S., O.P. acknowledge partial support from the INAF mini-grant 1.05.23.04.04. 
This project was partially funded also through the MSCA4Ukraine project from the European Union. Views and opinions expressed are however those of the authors only and do not necessarily reflect those of the European Union. Neither the European Union nor the MSCA4Ukraine Consortium as a whole nor any individual member institutions of the MSCA4Ukraine Consortium can be held responsible for them.
This work made use of the HPC system MEUSA, part of the Sistema Computazionale per l'Astrofisica Numerica (SCAN) of INAF-Osservatorio Astronomico di Palermo.
S.N. is supported by JSPS Grant-in-Aid Scientific Research (KAKENHI) (A), Grant Number JP25H00675 and JST ASPIRE Program “RIKEN-Berkeley mathematical quantum science initiative.
We acknowledge Andrea Damonte for his kind support on the MCMC chains.
\end{acknowledgements}

\bibliographystyle{aa}
\bibliography{references.bib}

\begin{appendix}
\begin{onecolumn}
\section{Observations}
\label{app:obs}

Tab. \ref{tab:obs} List of the \emph{Chandra} observations analyzed in this paper. Horizontal lines marks the different groups that we selected for the spectral analysis (see Sect. \ref{sec:results} and Appendix \ref{evolution}).

\begin{longtable}{lllrrr}
\caption{List of observations}\\
    \hline\hline
    Start Date & OBS ID & PI & RA & DEC & T$_{exp}$ (ks) \\
    \hline 
    2000-12-07 & 1967 &  McCray & 05 35 28.00 & -69 16 11.10 & 98.76 \\
    2001-12-12 & 2831 & Burrows & 05 35 28.00 & -69 16 11.10 & 49.41\\
	2002-05-15 & 2832 & Burrows & 05 35 28.00 & -69 16 11.10 & 44.26\\
	2002-12-31 & 3829 & Burrows & 05 35 28.00 & -69 16 11.10 & 49.01\\
	2003-07-08 & 3830 & Burrows & 05 35 28.00 & -69 16 11.10 & 45.31\\
	2004-01-02 & 4614 & Burrows & 05 35 28.00 & -69 16 11.10 & 46.49\\
	2004-07-22 & 4615 & Burrows & 05 35 28.00 & -69 16 11.10 & 48.83\\
	2004-08-26 & 4640 & McCray & 05 35 28.00 & -69 16 11.10 & 56.83 \\
	2004-08-27 & 5362 & McCray & 05 35 28.00 &-69 16 11.10	 & 67.42 \\
	2004-08-30 & 4641 & McCray & 05 35 28.00 & -69 16 11.10 & 72.48 \\
	2004-09-01 & 6099 & McCray & 05 35 28.00 & -69 16 11.10 & 48.61 \\
	2004-09-05 & 5363 & McCray & 05 35 28.00 & -69 16 11.10 & 43.45 \\
 \hline
	2005-01-09 & 5579 & Burrows & 05 35 28.00 & -69 16 11.10 & 31.87 \\
	2005-01-13 & 6178 & Burrows & 05 35 28.00 & -69 16 11.10 & 16.48 \\
	2005-07-11 & 5580 & Burrows & 05 35 28.00 & -69 16 11.10 & 23.75 \\
	2005-07-16 & 6345 & Burrows &05 35 28.00 & -69 16 11.10 & 20.99 \\
	2006-01-28 & 6668 & Burrows & 05 35 28.00 & -69 16 11.10 & 42.34 \\
	2006-07-27 & 6669 & Burrows & 05 35 28.00 & -69 16 11.10 & 36.45 \\
	2007-01-19 & 7636 & Burrows &05 35 28.00 & -69 16 11.10 & 33.51 \\
	2007-03-11 & 8523 & Canizares & 05 35 28.00 & -69 16 11.00 & 29.65 \\
	2007-03-12 & 8537 & Canizares & 05 35 28.00 & -69 16 11.00 & 12.73 \\
	2007-03-13 & 7588 & Canizares & 05 35 28.00 & -69 16 11.00 & 27.21 \\
	2007-03-18 & 8538 & Canizares & 05 35 28.00 & -69 16 11.00 & 20.69 \\
	2007-03-19 & 7589 & Canizares & 05 35 28.00 & -69 16 11.00 & 25.27 \\
	2007-03-20 & 8539 & Canizares & 05 35 28.00 & -69 16 11.00 & 24.78 \\
	2007-03-21 & 8542 & Canizares & 05 35 28.00 & -69 16 11.00 & 17.85 \\
	2007-03-24 & 8487 & Canizares & 05 35 28.00 & -69 16 11.00 & 28.67 \\
	2007-03-27 & 8543 & Canizares & 05 35 28.00 & -69 16 11.00 & 30.66 \\
	2007-03-28 & 8544 & Canizares & 05 35 28.00 & -69 16 11.00 & 19.12 \\
	2007-03-29 & 8488 & Canizares & 05 35 28.00 & -69 16 11.00 & 31.66 \\
	2007-03-31 & 8545 & Canizares & 05 35 28.00 & -69 16 11.00 & 20.46 \\
	2007-04-01 & 8546 & Canizares & 05 35 28.00 & -69 16 11.00 & 30.64 \\
	2007-04-17 & 7590 & Canizares & 05 35 28.00 & -69 16 11.00 & 35.55 \\
	2007-07-13 & 7637 & Burrows & 05 35 28.00 & -69 16 11.10 & 25.72 \\
	2007-09-04 & 9581 & McCray & 05 35 28.00 & -69 16 11.10	& 44.96 \\
	2007-09-05 & 9582 & McCray & 05 35 28.00 & -69 16 11.10 & 44.17 \\
	2007-09-07 & 9580 & McCray & 05 35 28.00 & -69 16 11.10 & 34.59 \\
	2007-09-09 & 7620 & McCray & 05 35 28.00 & -69 16 11.10 & 34.62 \\
	2007-09-11 & 7621 & McCray & 05 35 28.00 & -69 16 11.10	& 36.95 \\
	2007-09-12 & 9591 & McCray & 05 35 28.00 & -69 16 11.10 & 12.85 \\
	2007-09-12 & 9592 & McCray & 05 35 28.00 & -69 16 11.10 & 12.87 \\
	2007-09-14 & 9589 & McCray & 05 35 28.00 & -69 16 11.10 & 39.53 \\
	2007-09-16 & 9590 & McCray & 05 35 28.00 & -69 16 11.10 & 24.65 \\
	2008-07-01 & 9144 & Burrows & 05 35 28.00 & -69 16 11.10	& 42.03 \\
	2008-07-04 & 9143 & Burrows & 05 35 28.00 & -69 16 11.10 & 8.58 \\
	2009-01-05 & 10130 & Burrows & 05 35 28.00 & -69 16 11.10 & 6.02 \\
	2009-01-12 & 10852 & Burrows & 05 35 28.00 & -69 16 11.10 & 10.78 \\
	2009-01-13 & 10221 & Burrows & 05 35 28.00 & -69 16 11.10	& 18.73 \\
	2009-01-15 & 10853 & Burrows & 05 35 28.00 & -69 16 11.10 & 11.25 \\
	2009-01-17 & 10854 & Burrows & 05 35 28.00 & -69 16 11.10	& 11.99 \\
	2009-01-18 & 10855 & Burrows & 05 35 28.00 & -69 16 11.10 & 18.78 \\
	2009-07-06 & 10222 & Burrows & 05 35 28.00 & -69 16 11.10	& 23.47 \\
	2009-09-08 & 10926 & Burrows	& 05 35 28.00 & -69 16 11.10	& 33.83 \\
 \hline
	2010-03-17 & 12125 & Burrows & 05 35 28.00 & -69 16 11.10	& 18.15 \\
	2010-03-17 & 12126 & Burrows & 05 35 28.00 & -69 16 11.10 & 21.2 \\
	2010-03-28 & 11090 & Burrows & 05 35 28.00 & -69 16 11.10 & 24.56 \\
	2010-09-28 & 13131 & Burrows & 05 35 28.00 &	-69 16 11.10 & 26.48 \\
	2010-09-29 & 11091 & Burrows & 05 35 28.00 & -69 16 11.10 & 27.92 \\
	2011-03-01 & 12145 & Canizares & 05 35 28.00 & -69 16 11.00 & 51.21 \\
	2011-03-04 & 13238 & Canizares & 05 35 28.00 & -69 16 11.00 & 54.16 \\
	2011-03-06 & 13239 & Canizares & 05 35 28.00 & -69 16 11.00 & 46.74 \\
	2011-03-13 & 12146 & Canizares & 05 35 28.00 & -69 16 11.00 & 25.62 \\
	2011-03-25 & 12539 & Burrows	& 05 35 28.00 & -69 16 11.10 & 52.15 \\  
	2011-09-21 & 12540 & Burrows & 05 35 28.00 & -69 16 11.10 & 37.53 \\
	2011-09-22 & 14344 & Burrows & 05 35 28.00 & -69 16 11.10 & 11.59 \\
	2012-03-28 & 13735 & Burrows & 05 35 28.00 & -69 16 11.10 & 42.91 \\
	2012-04-01 & 14417 & Burrows & 05 35 28.00 & -69 16 11.10 & 26.94 \\
	2013-03-21 & 14697 & Burrows & 05 35 28.00 & -69 16 11.10 & 67.57 \\
	2013-09-28 & 14698 & Burrows & 05 35 28.00 & -69 16 11.10 & 68.53 \\
	2014-03-19 & 15809 & Burrows & 05 35 28.00 & -69 16 11.10 & 74.46 \\
	2014-09-17 & 17415 & Burrows & 05 35 28.00 & -69 16 11.10	& 19.37 \\
	2014-09-20 & 15810 & Burrows & 05 35 28.00 & -69 16 11.10 & 48.29 \\
    2015-09-17 & 16756 & Burrows & 05 35 28.00 &	-69 16 11.10 & 66.6 \\
    \hline
	2016-09-19 & 17899 & Burrows & 05 35 28.00 & -69 16 11.10 & 26.12 \\
	2016-09-23 & 19882 & Burrows & 05 35 28.00 & -69 16 11.10 & 41.08 \\
	2017-09-21 & 20793 & Burrows & 05 35 28.00 & -69 16 11.10	& 48.29 \\
	2017-09-23 & 19289 & Burrows & 05 35 28.00 & -69 16 11.10 & 18.9 \\
    2018-03-14 & 20927 & Park & 05 35 28.00 & -69 16 11.10 & 16.6 \\
    2018-03-15 & 21037 & Park & 05 35 28.00 & -69 16 11.10 & 29.42 \\
    2018-03-18 & 21038 & Park & 05 35 28.00 & -69 16 11.10 & 34.22 \\
    2018-03-19 & 20322 & Park & 05 35 28.00 & -69 16 11.10 & 15.16 \\
    2018-03-23 & 21042 & Park & 05 35 28.00 & -69 16 11.10 & 41.05 \\
    2018-03-25 & 21043 & Park & 05 35 28.00 & -69 16 11.10 & 28.83 \\
    2018-03-26 & 21044 & Park & 05 35 28.00 & -69 16 11.10 & 15.8 \\
    2018-03-27 & 20323 & Park & 05 35 28.00 & -69 16 11.10 & 27.09 \\
    2018-03-28 & 21049 & Park & 05 35 28.00 & -69 16 11.10 & 30.9 \\
    2018-03-29 & 21050 & Park & 05 35 28.00 & -69 16 11.10 & 17.76 \\
    2018-03-30 & 21051 & Park & 05 35 28.00 & -69 16 11.10 & 15.01 \\
    2018-03-31 & 21052 & Park & 05 35 28.00 & -69 16 11.10 & 30.12 \\
    2018-04-02 & 21053 & Park & 05 35 28.00 & -69 16 11.10 & 12.19 \\
    2018-09-15 & 20277 & Burrows	& 05 35 28.00 & -69 16 11.10 & 33.83 \\
    2018-09-16 & 21844 & Burrows & 05 35 28.00 & -69 16 11.10 & 33.83 \\
    2019-09-17 & 21304 & Burrows & 05 35 28.00 & -69 16 11.10 & 41.06 \\
    2019-09-18 & 22849 & Burrows & 05 35 28.00 & -69 16 11.10 & 41.06 \\
    2020-09-12 & 22425 & Burrows & 05 35 28.00 & -69 16 11.10 & 60.82 \\
    2020-09-17 & 24652 & Burrows & 05 35 28.00 & -69 16 11.10 & 29.33 \\
    2021-10-25 & 23534 & Burrows	& 05 35 28.00 & -69 16 11.10 & 29.00\\
    2021-10-27 & 24295 & Burrows &05 35 28.00 &-69 16 11.10	 & 30.00 \\
    2021-10-28 & 24654 & Burrows	& 05 35 28.00 & -69 16 11.10 & 30.00 \\
	\hline
	\hline
	\label{tab:obs}
\end{longtable}
\end{onecolumn}

\section{Flux temporal evolution}
\label{evolution}

\begin{figure*}[ht!]
{\includegraphics[width=0.33\columnwidth]{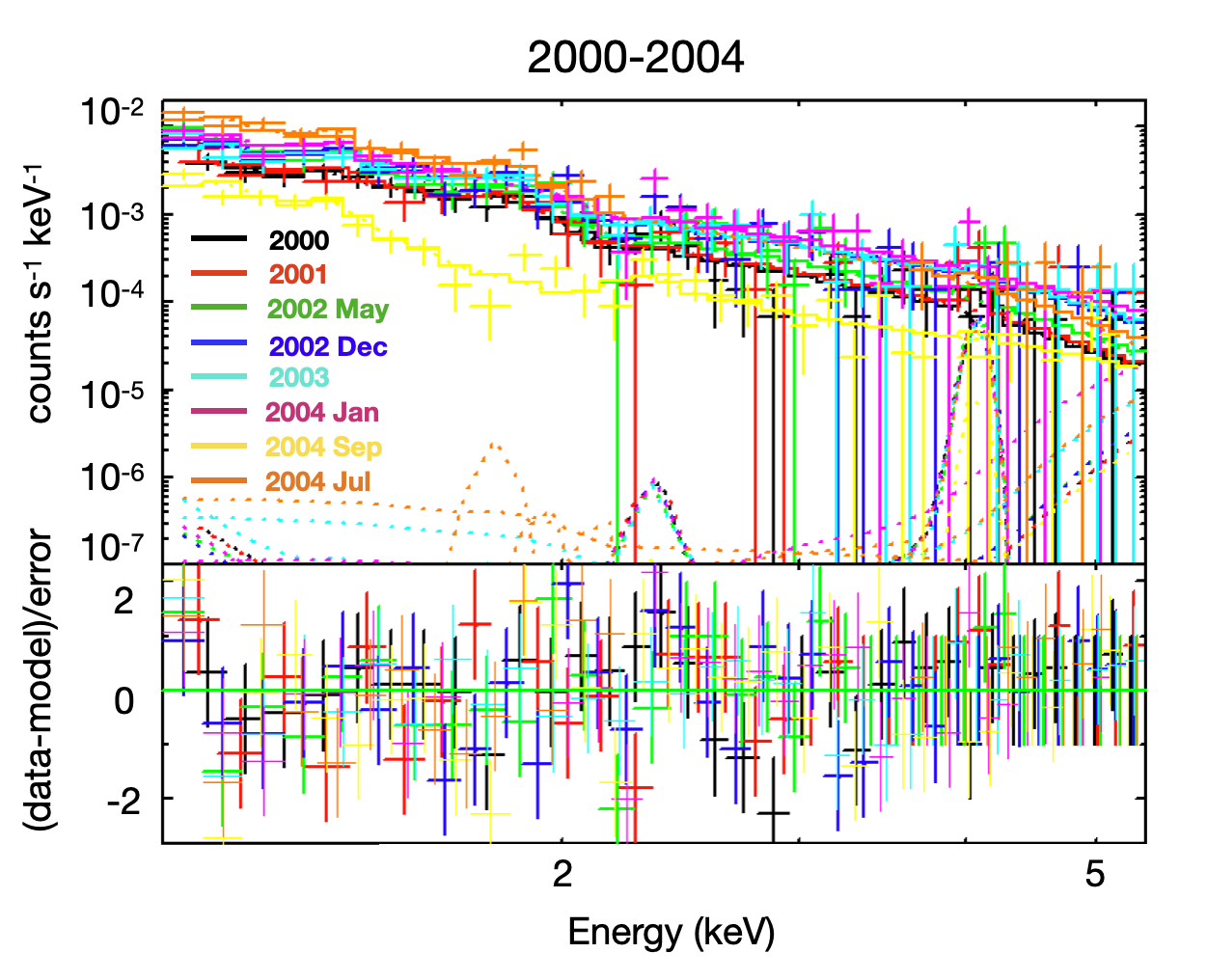}} 
{\includegraphics[width=0.33\columnwidth]{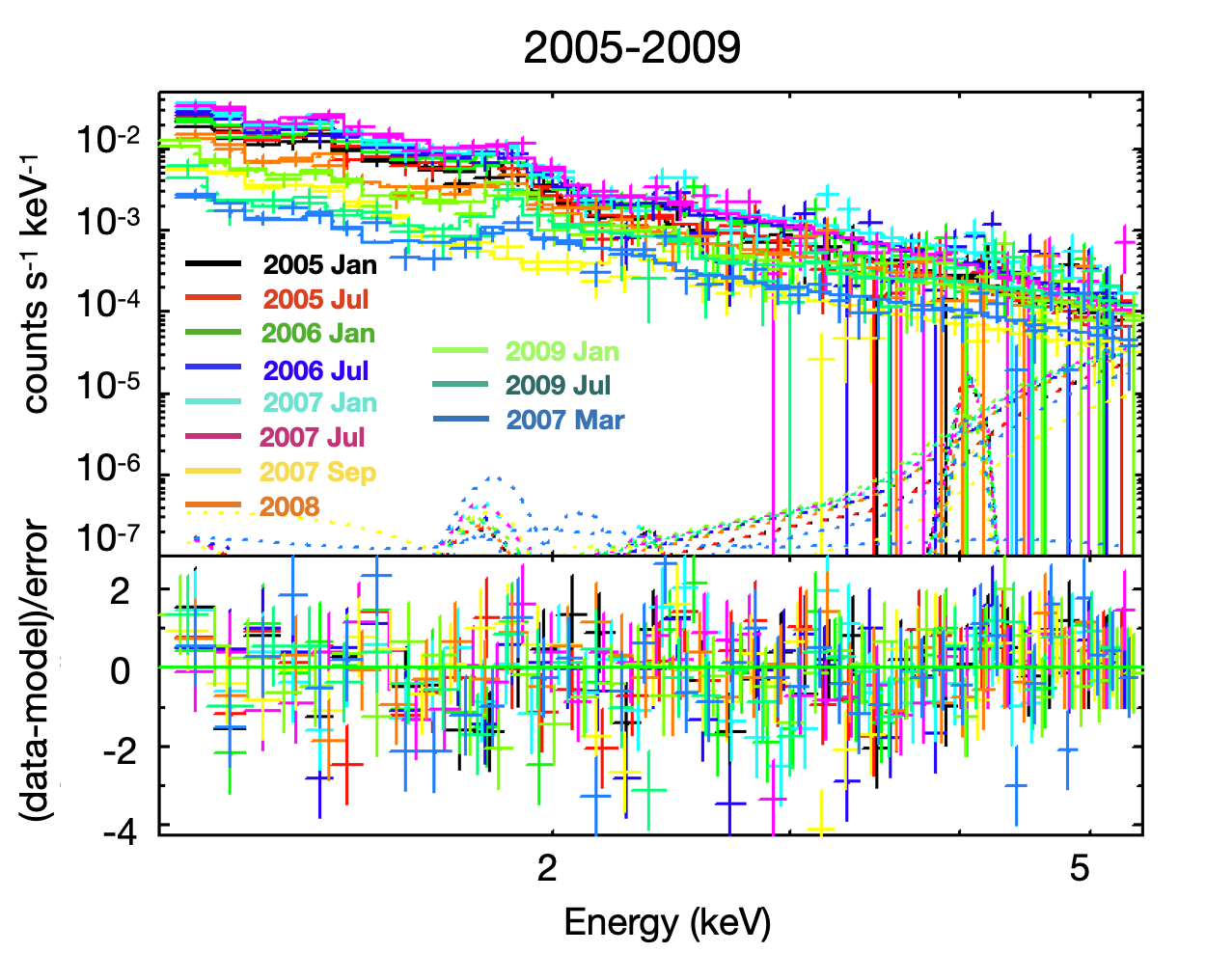}} 
{\includegraphics[width=0.33\columnwidth]{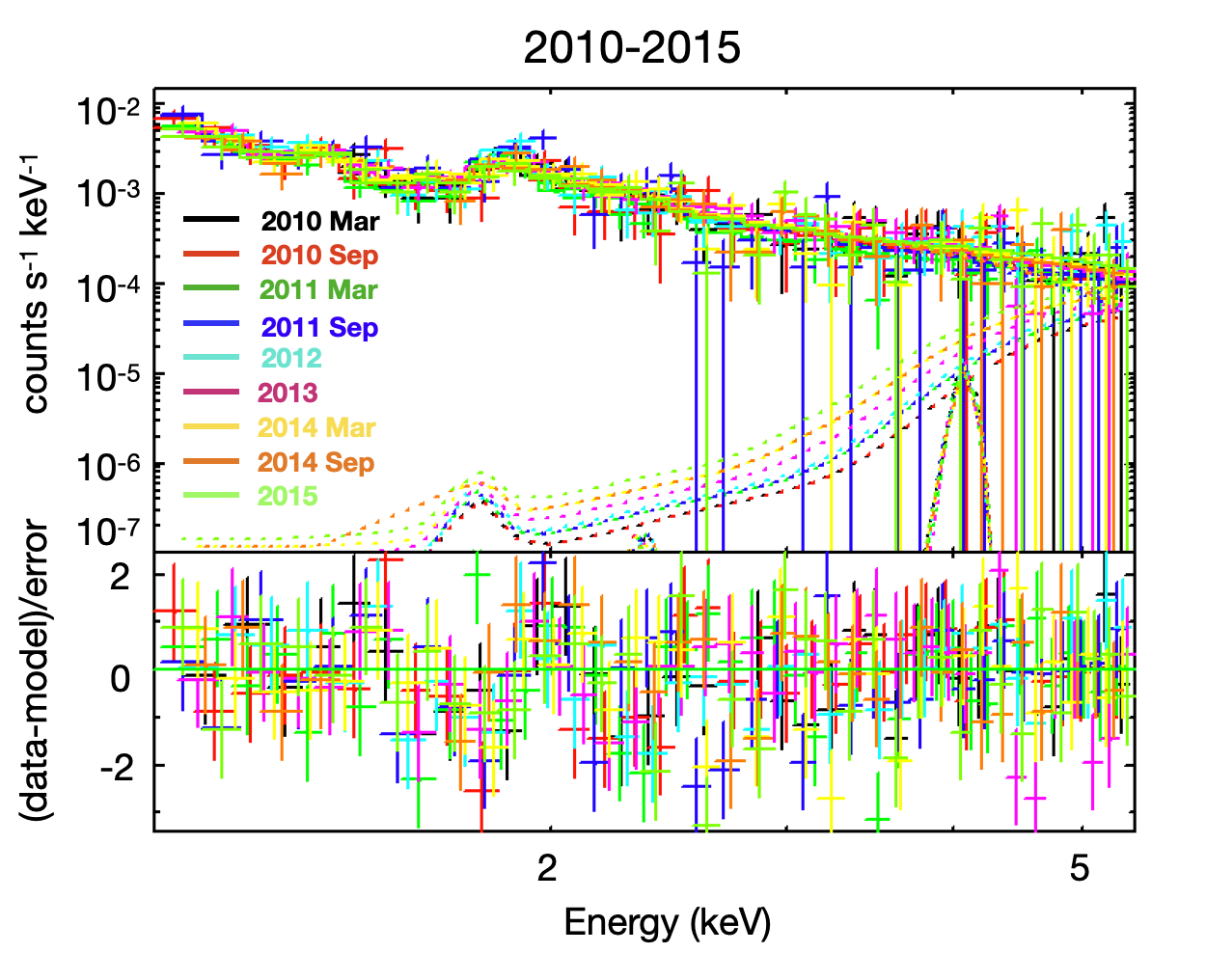}} 
	\caption{Chandra/ACIS spectra extracted from
the circular region shown in Fig. 1 for all observations performed between 2000 and 2004 (\emph{left}), 2005-2009 (\emph{center}), 2010-2015 (\emph{right}) with the corresponding best-fit model and residuals.}
	\label{fig:spectra}
\end{figure*}

Motivated by the detection of the $^{44}$Sc emission line in the 2016-2021 spectra, we repeated the spectral analysis described in Sect. \ref{sec:results} on earlier epochs, dividing the data into three additional groups, namely Group 1 for observations performed between 2000 and 2004, Group 2 for those in $2005-2009$ and Group 3 for $2010-2015$. Figure \ref{fig:spectra} shows the spectra of the different Groups. We measure the flux of the emission line centered at $4.076$ keV in all the Groups, the line flux (corrected for the PSF effects) being $F_1= 1.4^{+0.2}_{-0.3}\times 10^{-7}$ photons s$^{-1}$ cm$^{-2}$, $F_2< 3.9\times 10^{-8}$ photons s$^{-1}$ cm$^{-2}$, $F_3=7.3^{+2.2}_{-5.2}\times 10^{-8}$ photons s$^{-1}$ cm$^{-2}$ in Group 1, Group 2 and Group 3, respectively (error bars at 90\% confidence level).
As expected, the line flux in Groups $1-3$, is significantly lower than that in Group 4. This is likely the result of the cold ejecta being still optically thick at these epochs. However, discussing the time evolution of the line flux is beyond the scope of this paper, given its intrinsic complexity related to i) the complex effect of the absorption of cold ejecta, which monotonically (but non-linearly) decreases with time, ii) the expansion of the bright X-ray ring (which decreases the fraction of photons spilling from the ring into our extraction region) and iii) the increase of the hard X-ray flux ($3-8$ keV, see \citealt{rpz24}), which increases the contamination from the ring.

\begin{figure}[htb!]
\centering
\includegraphics[width=.48\textwidth]{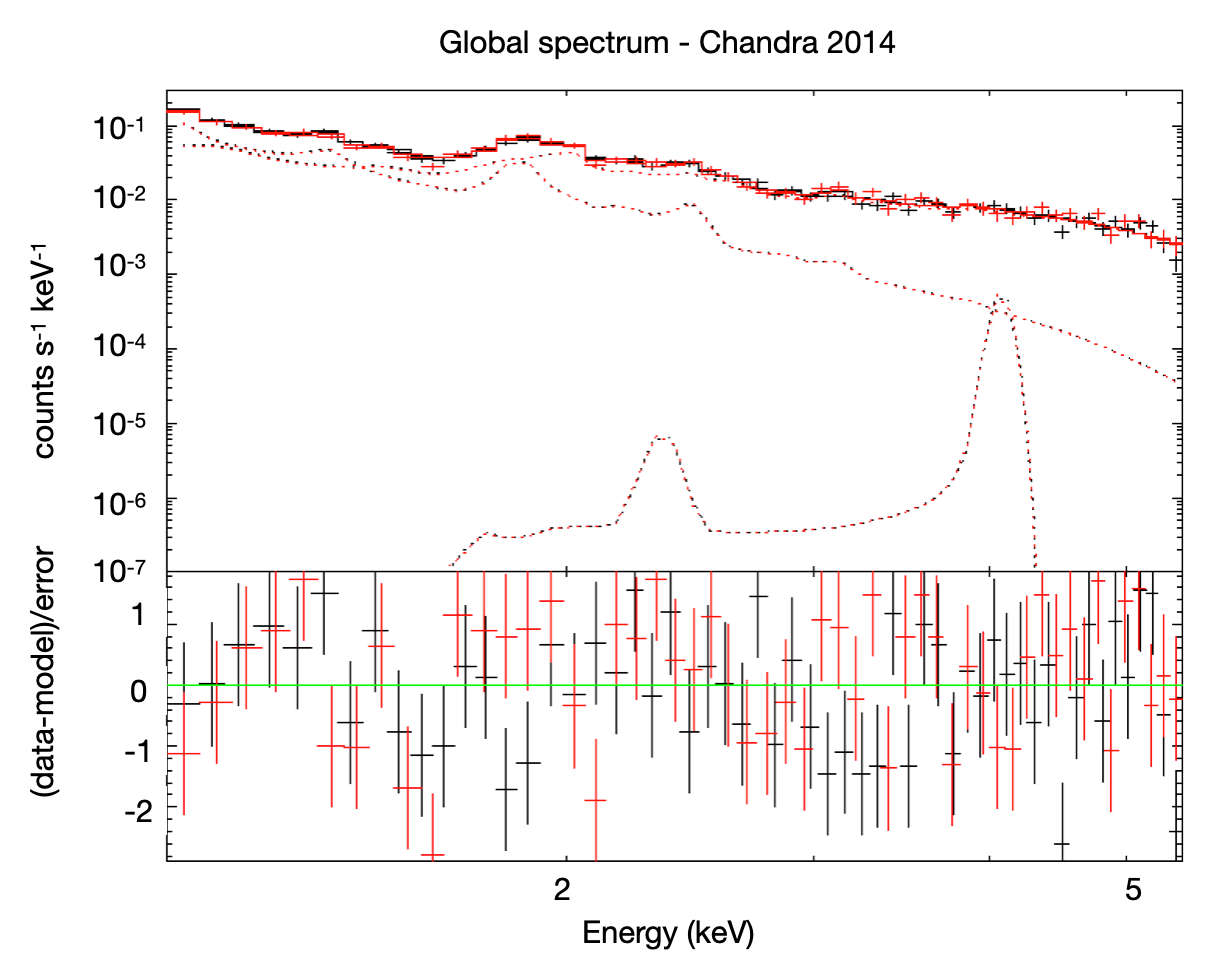}
\includegraphics[width=.48\textwidth]{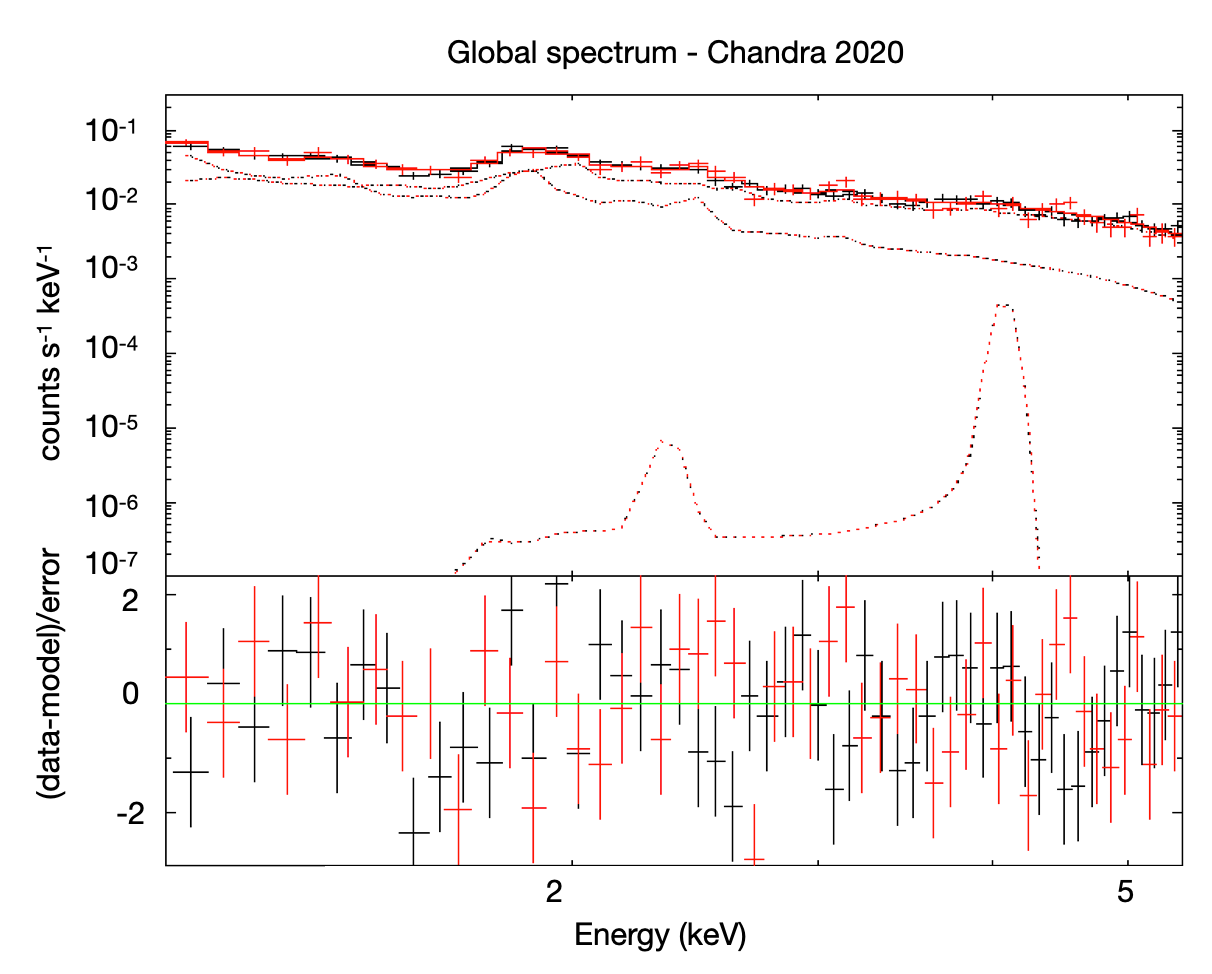}
	\caption{Spectra of the X-ray emission originating from a region including the ER of SN1987A at two different epochs (2014, \emph{left panel} and 2020, \emph{right panel}), with the corresponding best fit model and residuals.}
	\label{fig:87aring}
\end{figure}

We also checked that the line is not significantly detected in the spectra of the ring (as already shown by \citealt{Leising06} for the first \emph{Chandra} observations). As an example, we explored the spectra in 2014 and 2020 finding that the line flux is always compatible with 0 at the $68\%$ confidence level. Figure \ref{fig:87aring} shows that the line emission is always well below the continuum.

\section{Background spectrum}
\label{app:bkg}
\begin{figure}[htb!]
\includegraphics[width=0.48\textwidth]{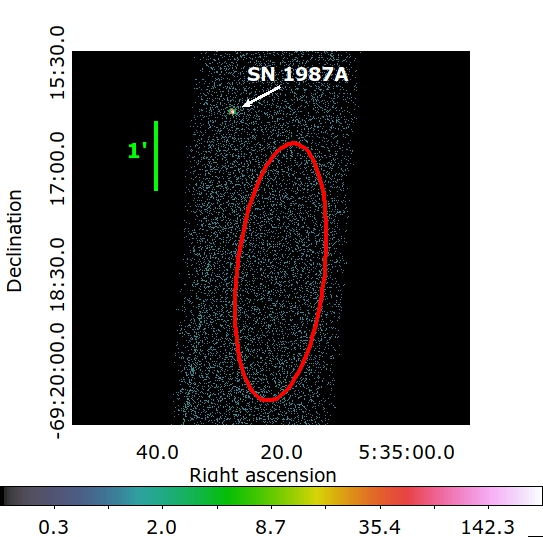}
\includegraphics[width=0.48\textwidth]{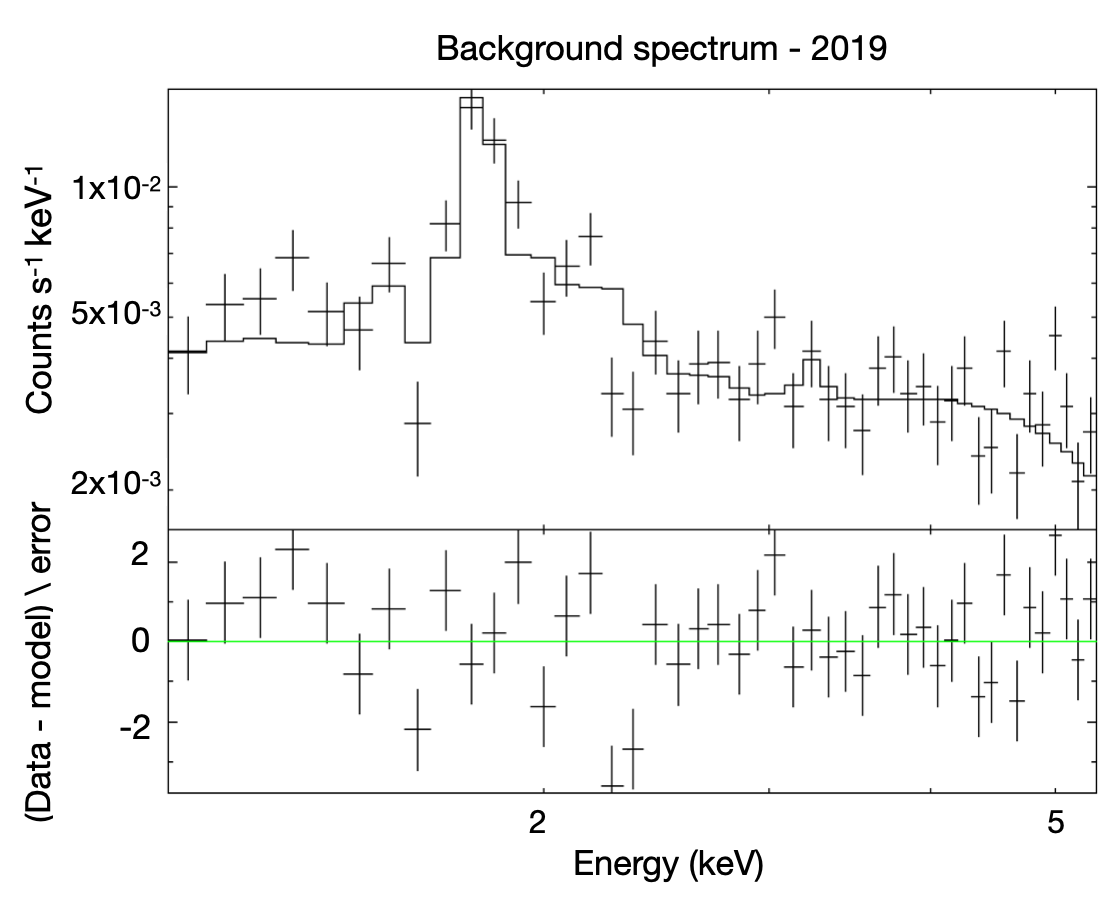}
	\caption{\emph{Left panel:} \emph{Chandra}/ACIS count map for Obs ID 21304 (same as in Fig. \ref{fig:img}) in log scale. The red ellipse shows the background extraction region. \emph{Right panel: }Spectrum extracted from the background region in the left panel with the corresponding best fit model (Eq. \ref{eq:bkg}) and residuals.}
	\label{fig:bkg_spectrum}
\end{figure}

Figure \ref{fig:bkg_spectrum} shows our choice of the background extraction region together with an example of the background spectrum, fitted with the following model:
\begin{equation}
    \textit{Bkg model} = \textit{const}*(pow + apec + gauss + gauss + gauss)
    \label{eq:bkg}
\end{equation}

In all of the background spectra the best fit model includes a power law taking into account the continuum, the model \texttt{apec} fitting the emission spectrum from collisionally-ionized diffuse gas , based on the database AtomDB version 3.09 \url{https://heasarc.gsfc.nasa.gov/xanadu/xspec/manual/XSmodelApec.html}, plus three emission lines centered at energies which are less than 3 keV.
\newpage
\section{Corner plot}
\label{app:corner}
MCMC corner plot for the simultaneous analysis of the spectra collected between 2016 and 2021. Each panel shows the correlation between two different free parameters, indicated as \texttt{parameter\_number}. In this case \texttt{norm\_24} is the flux associated with the $^{44}$Sc emission line. All the temperature, ionization times and normalization are associated with the \texttt{vnei} component fitted for each spectrum.

\begin{figure*}[ht!]
\includegraphics[width=\textwidth]{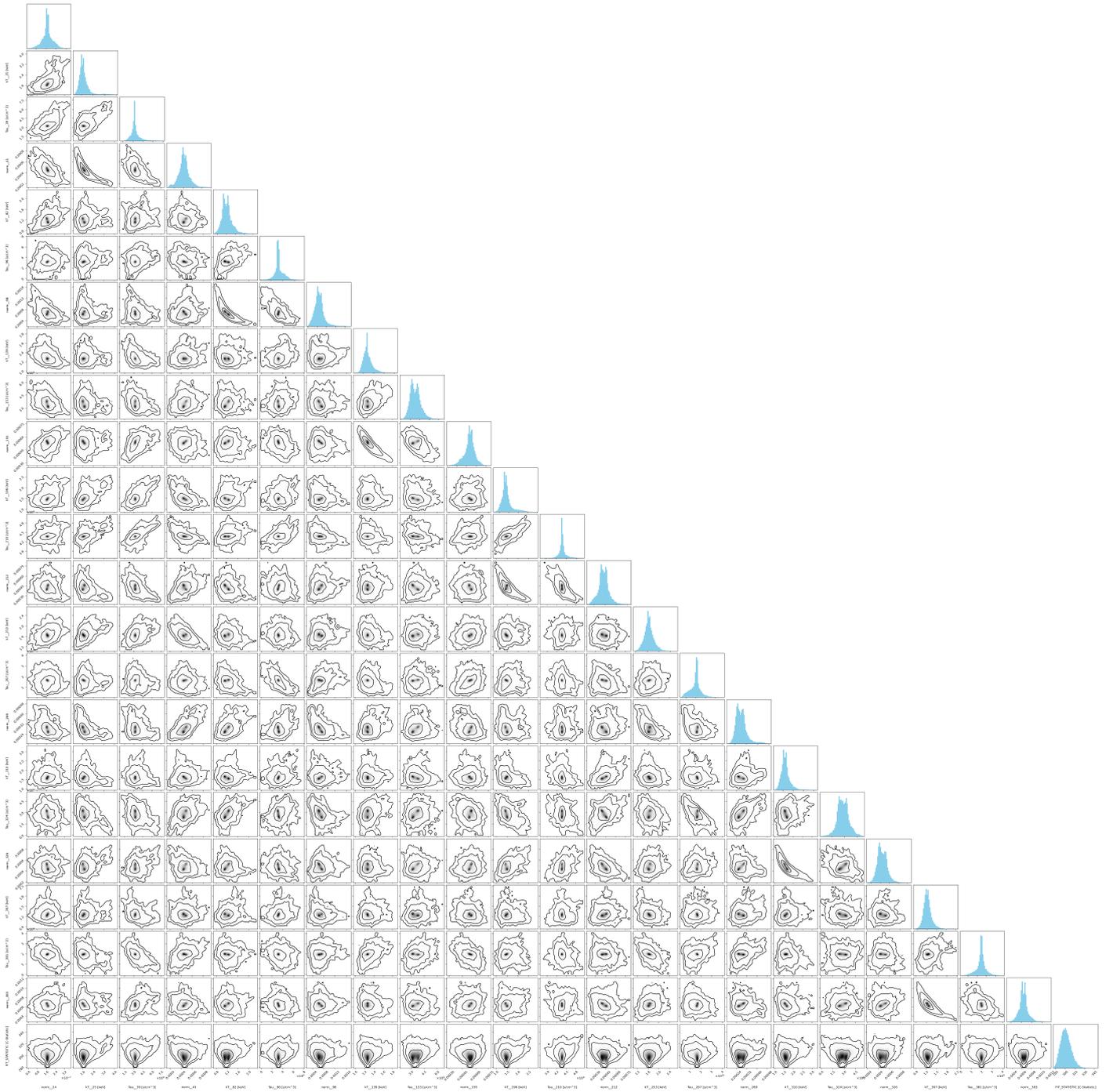}
	\caption{MCMC Corner plot for the simultaneous fit of 2016-2021 spectra (see Sect. \ref{sec:reduction} and Sect. \ref{sec:results}).}
	\label{fig:corner_plot_global}
\end{figure*}
\end{appendix}
\end{document}